\documentclass[runningheads]{llncs}
\usepackage[T1]{fontenc}
\usepackage{graphicx}

\usepackage{orcidlink}
\usepackage{fancyvrb}
\usepackage{listings}
\usepackage{subcaption}
\usepackage{svg} 
\usepackage{multirow}
\usepackage{array}

\DefineVerbatimEnvironment{CustomVerbatim}{Verbatim}{fontsize=\scriptsize }

\begin{document}
\title{Process Modeling With Large Language Models}
%

\authorrunning{H. Kourani et al.}
\author{
Humam Kourani\inst{1,2}\orcidlink{0000-0003-2375-2152} \and
Alessandro Berti\inst{1,2}\orcidlink{0000-0002-3279-4795} \and
Daniel Schuster\inst{1,2}\orcidlink{0000-0002-6512-9580} \and
Wil M. P. van der Aalst\inst{1,2}\orcidlink{0000-0002-0955-6940}
}
%
\institute{Fraunhofer Institute for Applied Information Technology FIT, Sankt Augustin, Germany \\
\and
RWTH Aachen University, Aachen, Germany \\
\email{\{humam.kourani, alessandro.berti, daniel.schuster, wil.van.der.aalst\}@fit.fraunhofer.de}
} 

\maketitle              
\begin{abstract}
In the realm of Business Process Management (BPM), process modeling plays a crucial role in translating complex process dynamics into comprehensible visual representations, facilitating the understanding, analysis, improvement, and automation of organizational processes. Traditional process modeling methods often require extensive expertise and can be time-consuming. This paper explores the integration of Large Language Models (LLMs) into process modeling to enhance the accessibility of process modeling, offering a more intuitive entry point for non-experts while augmenting the efficiency of experts. We propose a framework that leverages LLMs for the automated generation and iterative refinement of process models starting from textual descriptions. Our framework involves innovative prompting strategies for effective LLM utilization, along with a secure model generation protocol and an error-handling mechanism. Moreover, we instantiate a concrete system extending our framework. This system provides robust quality guarantees on the models generated and supports exporting them in standard modeling notations, such as the Business Process Modeling Notation (BPMN) and Petri nets. Preliminary results demonstrate the framework's ability to streamline process modeling tasks, underscoring the transformative potential of generative AI in the BPM field.

\keywords{Process Modeling \and Business Process Management \and Generative AI \and Large Language Models}
\end{abstract}

\section{Introduction}
\label{sec:introduction}
Process modeling is an essential aspect of Business Process Management (BPM), serving as a comprehensive toolkit for understanding, documenting, analyzing, and improving complex business operations. Business process modeling covers several formats -- from textual representations to visual diagrams and executable models -- thus facilitating a multifaceted approach to capturing organizational processes.

Business process modeling encompasses several key perspectives, each focusing on different process aspects. Traditionally, these perspectives include the \emph{control-flow perspective}, which outlines the flow of activities and their dependencies; the \emph{data perspective}, focusing on how data is generated, manipulated, and consumed throughout the process; the \emph{resource perspective}, detailing the human and system resources involved in the process execution; and the \emph{operational perspective}, which describes the operational rules and execution semantics. In this paper, we focus on enhancing the control-flow perspective of process modeling as the control-flow establishes the basic structure upon which the data, resource, and operational perspectives are built. 

Business process modeling traditionally involves extensive manual effort and deep knowledge of complex process modeling languages like BPMN (Business Process Model and Notation) \cite{DBLP:books/el/15/RosingWCM15} or Petri nets \cite{DBLP:journals/topnoc/HeeSW13a}. Additionally, process models often necessitate ongoing updates to reflect process changes. These challenges create significant barriers to entry for users without expertise in modeling languages, underscoring the need for new, streamlined process modeling methodologies.

The advent of Large Language Models (LLMs) such as GPT-4 \cite{DBLP:journals/corr/abs-2303-08774} and Gemini \cite{DBLP:journals/corr/abs-2312-11805} introduces a promising solution for enhancing the efficiency and accessibility of process modeling. Trained on diverse and extensive datasets, these LLMs show advanced capabilities in performing different tasks, ranging from coherent and contextually relevant text generation to solving complex problem-solving queries and generating executable code \cite{ijcai2021p612,DBLP:conf/hpec/VidanF23,DBLP:conf/iclr/ZhouMHPPCB23}. Their ability to understand and process complex textual information in natural language makes LLMs particularly well-suited for process modeling and other tasks that require generating and refining structured outputs directly from textual descriptions. Therefore, leveraging LLMs in process modeling heralds a transformative shift, potentially reducing the dependence on manual effort and specialized knowledge.

Our paper introduces a novel framework that utilizes the power of LLMs to automate the generation of process models. It incorporates advanced techniques in prompt engineering, error handling, and code generation to transform textual process descriptions into process models illustrating the described processes. Additionally, our framework features an interactive feedback loop, allowing for refining the generated models based on the user's feedback. To demonstrate the feasibility and practical application of our framework, we implement a concrete system that instantiates it. This system leverages the Partially Ordered Workflow Language (POWL) \cite{powl} for the intermediate process representation, providing robust guarantees on the quality of the generated models. The generated POWL models can then be viewed and exported in standard modeling notations such as BPMN and Petri nets. We integrate the implemented system with state-of-the-art LLMs, showing the framework's ability to streamline process modeling and underscoring the potential of generative AI in revolutionizing BPM.

The remainder of this paper is structured as follows. In \autoref{sec:background}, we discuss related work. \autoref{sec:methodology} outlines our LLM-based process modeling framework. \autoref{sec:ev} evaluates the integration of our framework with state-of-the-art LLMs. In \autoref{sec:limit}, we discuss the limitations of our framework and propose ideas for future work. Finally, \autoref{sec:conclusion} concludes the paper.

\section{Related Work}
\label{sec:background}
An overview of various approaches for extracting process information from text is provided in \cite{DBLP:conf/aiia/BellanDG20}. While \cite{DBLP:journals/jucs/GoncalvesSB11} leverages Natural Language Processing (NLP) and text mining techniques to derive process models from text, \cite{DBLP:conf/caise/FriedrichMP11} combines NLP with computational linguistics techniques to generate BPMN models. In \cite{DBLP:journals/jksucis/SholiqSA22}, the authors employ NLP techniques to extract structured relationship representations, termed \emph{fact types}, from text, and the derived fact types are subsequently transformed into BPMN components. The BPMN Sketch Miner \cite{DBLP:conf/models/IvanchikjSP20} leverages process mining \cite{DBLP:books/daglib/0027363} to generate BPMN models starting from text in a \emph{domain-specific language}. Commercial vendors are integrating AI into process modeling, e.g., Process Talks (\url{https://processtalks.com}) provides an AI-powered system for creating process models starting from textual descriptions.

The integration of LLMs in BPM has been explored recently. Several studies \cite{DBLP:conf/bpmds/BuschRSL23,DBLP:conf/bpm/VidgofBM23} delve into the potential applications and challenges of employing LLMs for BPM tasks. In \cite{20.500.12116/43782}, limitations of using GPT-4 in conceptual modeling are discussed. LLMs are evaluated on various process mining tasks in \cite{DBLP:conf/bpm/Berti0A23}. The proposed approach in \cite{DBLP:journals/aai/ChenL22b} employs BERT \cite{DBLP:conf/naacl/DevlinCLT19} for the classification and analysis of process execution logs, aiming to improve process monitoring and anomaly detection. In \cite{DBLP:conf/bpm/KlievtsovaBKMR23}, the authors explore the novel concept of conversational modeling with LLMs, proposing a method for generating process models through dialogue-based interactions. The paper \cite{DBLP:conf/bpm/GrohsAER23} demonstrates the capability of LLMs to translate textual descriptions into procedural and declarative process model constraints. Finally, \cite{DBLP:journals/emisaij/FillFK23} investigate the broader implications of LLMs in conceptual modeling, suggesting potential applications beyond traditional BPM tasks.

\section{LLM-Based Process Modeling Framework}
\label{sec:methodology}
In this section, we detail our framework that leverages the power of LLMs for generating process models starting from process descriptions in natural language.

\subsection{Framework Overview}
\begin{figure}[!t]
    \centering        
    \includegraphics[width=\textwidth]{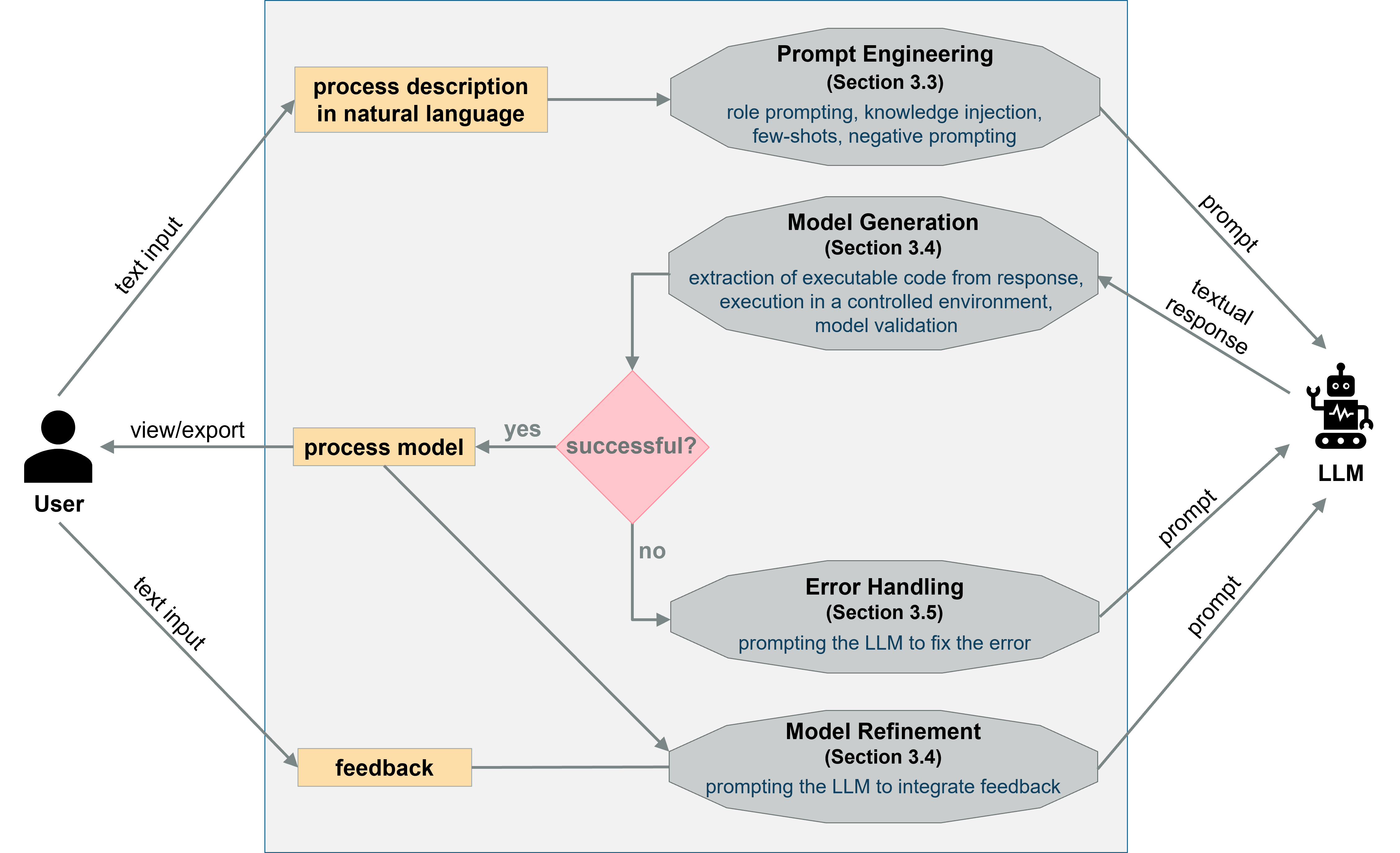}
    \caption{LLM-based process modeling framework.}
     \label{fig:architecture}
 \end{figure}

\autoref{fig:architecture} provides a schematic overview of our proposed framework. First, users input a textual description of a process in natural language. Upon receiving the textual description, we incorporate additional information to craft a comprehensive prompt (the employed prompt engineering techniques are detailed in \autoref{sec:prompt}). This prompt is designed to guide the LLM to generate executable code that can be used for the generation of process models (the selection of the modeling language used for process representation is discussed in \autoref{sec:powl}). The code generation step leverages a set of functions we designed to facilitate the creation of process models. After the prompt is generated, it is dispatched to the LLM. Note that the framework is independent of the selected LLM; it can be integrated with any advanced LLM that offers a large context window and code generation capabilities. After receiving the LLM's response, we extract the code snipped from the response and try to execute it (cf. \autoref{sec:gen}). In instances where the code extraction or execution encounters errors, we employ an error-handling mechanism that involves sending a refined prompt back to the LLM, exploiting LLMs' self-refinement capabilities to fix the error (cf. \autoref{sec:error}). Upon successful code execution and process model generation, users can view or export the model using established process modeling notations, such as BPMN and Petri nets. Moreover, the framework incorporates an interactive feedback loop. Users can provide feedback on the generated model, which is subsequently integrated into the model. This feature enables the continuous optimization and refinement of the generated process model.

\subsection{Process Representation}\label{sec:powl}

To better explain the different stages within our framework, we instantiate a concrete system that utilizes the Partially Ordered Workflow Language (POWL) \cite{powl} for intermediate process representation. The foundational principles of our framework allow for the integration with other modeling languages tailored to the requirements of process modeling. In this section, we motivate our choice of the POWL language.

Our aim is to generate process models in standard notations familiar to most professionals in the business process management field, such as BPMN and Petri nets. However, such modeling languages are complex with a high potential for quality issues. For example, it is possible to generate Petri nets or BPMN models with dead parts that can never be reached. Therefore, the concept of \emph{soundness} is introduced, and many approaches for the automated discovery of process models use process modeling languages that guarantee soundness (e.g., \cite{DBLP:conf/icpm/KouraniSA23,DBLP:series/lnbip/Leemans22}). Our system for process modeling employs POWL for intermediate process representation, and the generated POWL models are then transformed into BPMN or Petri nets. A POWL model is a partially ordered graph extended with control-flow operators for modeling choice and loop structures. POWL represents a subclass of Petri nets that allow for the generation of hierarchical models where sub-models are combined to generate larger ones. 

We have selected POWL as an intermediate process representation due to the following reasons: 
\begin{itemize}
    \item \textit{Soundness Guarantees:} Unlike BPMN models or Petri nets, POWL inherently ensures soundness.\textbf{}
    \item \textit{Simplicity:} POWL's hierarchical nature enables a simplified model generation by recursively generating models and combining them into larger ones. Moreover, POWL allows for combining sub-models as partial orders under the assumption that tasks are inherently parallel unless explicitly defined otherwise. This assumption of concurrent task execution mirrors the dynamics of numerous real-life processes, and it facilitates the generation of process models as the order of concurrent elements does not need to be specified.
    \item \textit{Expressive Power:} While POWL and process trees \cite{DBLP:series/lnbip/Leemans22} both guarantee soundness, POWL supports a broader range of process structures \cite{powl}. POWL allows for modeling intricate, non-hierarchical dependencies while maintaining the quality guarantees of hierarchical process modeling languages. 
\end{itemize}

\subsection{Prompt Engineering}\label{sec:prompt}
This section discusses the prompting strategies we employ to effectively utilize LLMs for process modeling. We guide the LLM toward a precise comprehension of the provided process descriptions and the subsequent generation of the targeted process models. These strategies aim to leverage the inherent capabilities of LLMs without the need for retraining or adjustments.

The following prompting strategies are employed within our process modeling framework:

\begin{itemize}
\item \textit{Role Prompting:}
This strategy involves assigning a specific role to the LLM to guide its responses or behavior in a particular direction \cite{DBLP:journals/corr/abs-2305-14688}. We implemented role prompting by instructing the LLM to act as an expert in process modeling, familiar with common process constructs. The LLM is also tasked to assume the role of a process owner and to use its expertise in the context of the process to fill in any gaps in the provided process description.


\item \textit{Knowledge Injection:} This strategy involves providing the LLM with new, specific information or context that it may not have been exposed to during its initial training \cite{DBLP:conf/esws/MartinoIT23}. We provide comprehensive knowledge about POWL, offering detailed insights into its hierarchical structure and the semantics of the different POWL components. Moreover, our framework leverages LLM capabilities in generating executable code \cite{DBLP:conf/hpec/VidanF23} by instructing the LLM to generate Python code that utilizes a predefined set of functions we designed for the safe generation of POWL models. We provide a detailed explanation of these predefined methods and how they can be used to generate POWL models. \autoref{lst:injection} illustrates the injected knowledge about POWL.

\begin{lstlisting}[caption={Injecting the LLM with knowledge about POWL. Lines that extend beyond the displayed text are abbreviated with ``...'' to keep it compact.}, frame=single, label={lst:injection}, float, floatplacement='h', basicstyle=\scriptsize\ttfamily]
Use the following knowledge about the POWL modeling language: A POWL ...
Provide the Python code that recursively generates a POWL model. Save the ...
Assume the class ModelGenerator is properly implemented and can be ... 
ModelGenerator provides the functions described below:
 - activity(label) generates an activity. It takes 1 string argument, ...
 - xor(*args) takes n >= 2 arguments, which are the submodels. Use it to ...
 - loop(do, redo) takes 2 arguments, which are the do and redo parts. Use ...
 - partial_order(dependencies) takes 1 argument, which is a list of ...
Note: for any powl model, you can call powl.copy() to create another ...
\end{lstlisting}

\item \textit{Few-Shots Learning:}
This method involves training the LLM on solving the task by providing several example pairs of input and expected output \cite{DBLP:conf/nips/BrownMRSKDNSSAA20}. This enhances the LLM's ability to generate POWL models starting from process descriptions. For instance, \autoref{lst:few-shots} shows one the examples we use for training. This example shows how to generate a POWL model for the bicycle manufacturing process from \cite{DBLP:conf/bpm/BellanADGP22}.

\item \textit{Negative Prompting:} Negative prompting refers to instructing the LLM by specifying what it should avoid in its response \cite{DBLP:journals/corr/abs-2305-16807}. We implement negative prompting by instructing the LLM to avoid common errors that can occur using our predefined methods for generating POWL models (e.g., trying to generate partial orders that violate irreflexivity). Moreover, we extend our few-shot demonstrations with common mistakes that should be avoided during the construction of each process. For example, a common mistake for the bicycle manufacturing process (cf. \autoref{lst:few-shots}) is to create a local choice between two activities ``reject order'' and ``accept order'' instead of modeling a choice between the complete paths that are taken in each case.
\end{itemize}

\subsection{Model Generation and Refinement}\label{sec:gen}
After receiving the LLM's response, the Python code snippet is extracted from the response, which might also include additional text (e.g., intermediate reasoning steps). If the code extraction is successful, then the extracted code is executed to generate the model. Executing code generated by an LLM involves multiple considerations to handle security risks and invalid results. The following strategies are implemented to ensure a safe environment for producing valid process models:

\clearpage
\begin{lstlisting}[caption={POWL model generation example used for few-shots learning, extended with instructions to avoid common errors. Lines that extend beyond the displayed text are abbreviated with ``...'' to keep it compact.}, frame=single, label={lst:few-shots},  basicstyle=\scriptsize\ttfamily]
Process description for example 1:
A small company manufactures customized bicycles. Whenever the sales ...

Process model for example 1:
```python
from utils.model_generation import ModelGenerator
gen = ModelGenerator()
create_process = gen.activity('Create process instance')
reject_order = gen.activity('Reject order')
accept_order = gen.activity('Accept order')
inform = gen.activity('Inform storehouse and engineering department')
process_part_list = gen.activity('Process part list')
check_part = gen.activity('Check required quantity of the part')
reserve = gen.activity('Reserve part')
back_order = gen.activity('Back-order part')
prepare_assembly = gen.activity('Prepare bicycle assembly')
assemble_bicycle = gen.activity('Assemble bicycle')
ship_bicycle = gen.activity('Ship bicycle')
finish_process = gen.activity('Finish process instance')

check_reserve = gen.xor(reserve, back_order)

single_part = gen.partial_order(dependencies=[(check_part, check_reserve)])
part_loop = gen.loop(do=single_part, redo=None)

accept_poset = gen.partial_order(
    dependencies=[(accept_order, inform),
                  (inform, process_part_list),
                  (inform, prepare_assembly),
                  (process_part_list, part_loop),
                  (part_loop, assemble_bicycle),
                  (prepare_assembly, assemble_bicycle),
                  (assemble_bicycle, ship_bicycle)])

choice_accept_reject = gen.xor(accept_poset, reject_order)

final_model = gen.partial_order(
    dependencies=[(create_process, choice_accept_reject),
                  (choice_accept_reject, finish_process)])
```

Common errors to avoid for example 1:
creating a local choice between 'reject_order' and 'accept_order' instead ...
\end{lstlisting}

\begin{itemize}
    \item In order to eliminate the risk of executing unsafe code, we restrict the LLM to use the predefined functions we designed for the generation of POWL models. We employ a strict process to verify that the code strictly complies with the prompted coding guidelines, explicitly excluding the use of external libraries or constructs that may pose security threats. 
    \item We define validation rules to ensure that the code generates models that align with the POWL specifications and requirements. For example, we validate that all partial orders within the generated model adhere to the transitivity and irreflexivity requirements.
\end{itemize}

Our framework converts the generated POWL models into Petri nets and BPMN models. It offers functionalities for displaying and exporting the models in these established notations, which are widely acknowledged within the business process management community.

\paragraph{Refinement Loop.} The framework supports model refinement based on user feedback. Users can provide comments on the generated model, and we prompt the LLM to update the model accordingly. These feedback prompts are sent along with the full conversation history. This interactive approach ensures continual improvement and customization of the models.

\subsection{Error Handling}\label{sec:error}

Despite their advanced coding capabilities, LLMs do not always generate error-free code. We employ a robust error-handling mechanism tailored to mitigate potential inaccuracies and ensure the reliability of the generated process models. 

Recognizing the variability in the severity and implications of errors, we categorize them into two distinct groups:
\begin{itemize}
    \item \textit{Critical Errors:} This category covers errors that significantly disrupt the system's functionality or compromise security. These encompass execution failures, security risks, and major model validation violations. Given their potential impact, critical errors necessitate decisive action and cannot be overlooked.

    \item Adjustable Errors: This category includes errors related to the model's qualitative aspects, such as the reuse of submodels within the same POWL model. Although they affect the model's precision or quality, adjustable errors are considered less critical. They can be adjusted automatically, allowing for a degree of flexibility in their resolution. For example, the error of reusing submodels within the same POWL model can be automatically resolved by creating copies of the reused models. However, such intervention is approached with caution to prevent significant deviations from the behavior of the intended process.
\end{itemize}

Our framework incorporates an iterative error-handling loop, engaging the LLM in resolving identified errors. A new prompt that details the error and requests the LLM to address it, along with the conversation history, are submitted to the LLM. This iterative cycle facilitates dynamic correction, leveraging the LLM's capabilities to refine and improve the generated code. 

For critical errors, the system persistently seeks resolution via the LLM up to a predetermined number of allowed attempts. If the LLM fails to fix the error after the allowed number of attempts, the system terminates the process and marks the model generation as unsuccessful. In cases of adjustable errors, the system initially attempts correction through iterative engagement with the LLM. If the LLM fails to resolve adjustable errors after several attempts, then the system automatically resolves these errors.

\section{Evaluation}\label{sec:ev}
In this section, we evaluate our LLM-based process modeling framework. We integrate the implemented system with state-of-the-art LLMs to demonstrate the feasibility and practical application of our framework. 

\subsection*{Research Questions}
We structure the evaluation around the following two research questions:

\begin{itemize}
    \item Q1: How does our framework perform when integrated with state-of-the-art LLMs?
    \item Q2: How does our framework's performance compare to other LLM-based process modeling systems?
\end{itemize}

Q1 aims to investigate the capability of our framework to leverage the latest advancements in LLM technology. We used two state-of-the-art LLMs: GPT-4 and Gemini. We focus on the framework's ability to generate accurate and optimized process models based on initial descriptions and through the iterative feedback loop. The assessment considers the quality of the generated models, the efficiency in handling errors, and the effectiveness of integrating user feedback.

Q2 aims to compare our process modeling framework with other existing approaches. Notably, we found no LLM-based techniques that directly produce process models in the literature. The closest related work is the framework proposed in \cite{DBLP:conf/bpm/GrohsAER23}, which utilizes LLMs to transform process descriptions in natural language into \emph{textual abstractions} in a pre-defined notation that captures BPMN base components. We use this framework to generate process models with GPT-4, and we manually transform the generated textual abstractions into BPMN models for comparative analysis. We refer to this approach as the TA (Textual Abstraction) framework throughout this paper. 

\subsection*{Setup}
We implemented an integration of our system with GPT-4 as a web application available at \url{https://promoai.streamlit.app/}. As Gemini APIs are not available in Germany, direct integration of our framework with Gemini is not feasible; we used the web interface (\url{https://gemini.google.com/app}), and we manually transferred the prompts and responses between Gemini and our framework.

Throughout our experiments, we set a threshold of two iterations for handling adjustable errors through interaction with the LLM before automatically resolving them, and we set a threshold of five iterations for critical errors before terminating the process and marking the model generation as unsuccessful. Each experiment was repeated three times to account for the non-deterministic nature of LLM results.

\begin{table}[!t]
\caption{Process description and feedback comments for the online shop process.}
\centering
\scriptsize
\begin{tabular}{|p{0.15\linewidth}|p{0.82\linewidth}|}
\hline
Initial Process Description & Consider a process for purchasing items from an online shop. The user starts an order by logging in to their account. Then, the user simultaneously selects the items to purchase and sets a payment method. Afterward, the user either pays or completes an installment agreement. After selecting the items, the user chooses between multiple options for a free reward. Since the reward value depends on the purchase value, this step is done after selecting the items, but it is independent of the payment activities. Finally, the items are delivered. The user has the right to return items for exchange. Every time items are returned, a new delivery is made. \\ \hline
1st Feedback & Model the item selection using an activity ``Add Items'' that can be repeated.  \\ \hline
2nd Feedback & The user may skip the reward selection. \\ \hline
\end{tabular}
\label{tab:online_shop}
\end{table}

\begin{table}[!t]
\caption{Process description and feedback comments for the hotel process.}
\centering
\scriptsize
\begin{tabular}{|p{0.15\linewidth}|p{0.82\linewidth}|}
\hline
Initial Process Description & The Evanstonian is an upscale independent hotel. When a guest calls room service at The Evanstonian, the room-service manager takes down the order. She then submits an order ticket to the kitchen to begin preparing the food. She also gives an order to the sommelier (i.e., the wine waiter) to fetch wine from the cellar and to prepare any other alcoholic beverages. Eighty percent of room-service orders include wine or some other alcoholic beverage. Finally, she assigns the order to the waiter. While the kitchen and the sommelier are doing their tasks, the waiter readies a cart ( i.e., puts a tablecloth on the cart and gathers silverware ). The waiter is also responsible for nonalcoholic drinks. Once the food, wine, and cart are ready, the waiter delivers it to the guest’s room. After returning to the room-service station, the waiter debits the guest’s account. The waiter may wait to do the billing if he has another order to prepare or deliver. \\ \hline
1st Feedback & Include an activity ``prepare food''.  \\ \hline
2nd Feedback & The guest may or may not tip the waiter after receiving the order.  \\ \hline
\end{tabular}
\label{tab:hotel}
\end{table}

We used two processes for our evaluation: the process described in \cite{powl} for handling orders in an online shop and the hotel service process from the PET data set \cite{DBLP:conf/bpm/BellanADGP22}. The process descriptions and feedback comments we used are reported in \autoref{tab:online_shop} for the online shop process and in \autoref{tab:hotel} for the hotel service process. Note that we incorporated the feedback comments into the process description when applying the TA approach.

\begin{table}[!t]
\centering
\caption{Number of error-handling iterations needed for the initial model generation and feedback integration for each process. We use * to indicate that adjustable errors were resolved automatically, not through interaction with the LLM. We use - to mark the cases where the model generation was unsuccessful after five error-handling iterations.}
\label{tab:iterations_llm_grouped}
\begin{tabular}{| c | c | ccc | ccc |}
\hline
\textbf{Process} & \textbf{Step} & \multicolumn{3}{c|}{\textbf{GPT-4}} & \multicolumn{3}{c|}{\textbf{Gemini}} \\
& & \textbf{run 1} & \textbf{run 2} & \textbf{run 3} & \textbf{run 1} & \textbf{run 2} & \textbf{run 3} \\
\hline
\multirow{3}{*}{Online Shop} & Initial Model & 2 & 1 & 2 & 2* & 2* & 2* \\
 & 1st Feedback    & 0 & 0 & 0 & 0  & -  & -  \\
 & 2nd Feedback   & 0 & 0 & 0 & -  & -  & -  \\
 \hline
 \multirow{3}{*} {Hotel} & Initial Model & 2 & 2* & 1 & 5* & - & 3* \\
 & 1st Feedback    & 0 & 2* & 0 & 2*  & -  & 2* \\
 & 2nd Feedback   & 0 & 2* & 0 & - & - & 2* \\
 \hline
\end{tabular}
\end{table}

\begin{figure}[!t]
    \centering   
    \begin{subfigure}[b]{\textwidth}
        \includegraphics[width=\textwidth]{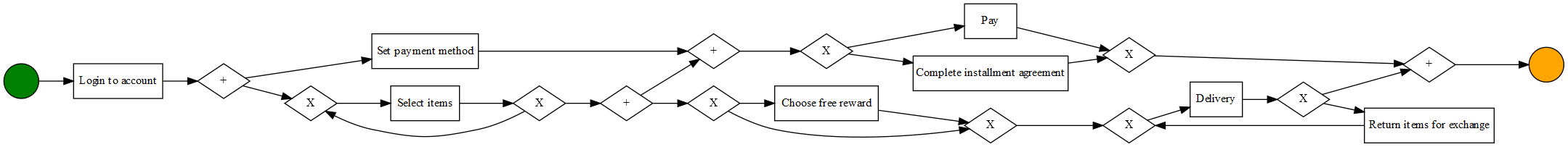}
        \caption{BPMN generated by GPT-4 using our system.}
    \end{subfigure}
    
    \begin{subfigure}[b]{\textwidth}
        \includegraphics[width=\textwidth]{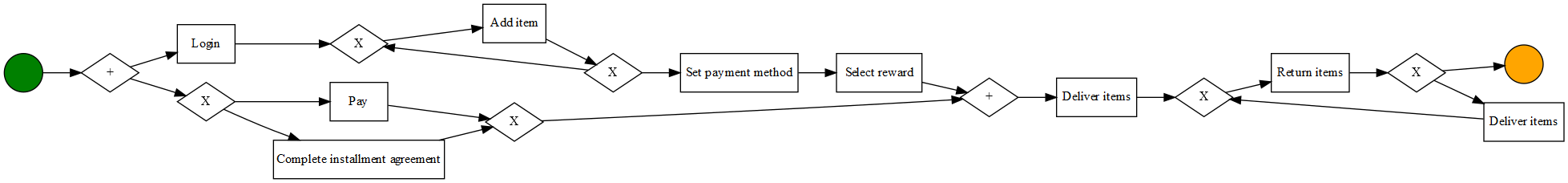}
        \caption{BPMN generated by Gemini using our system.}
    \end{subfigure}
    
    \begin{subfigure}[b]{\textwidth}
        \includegraphics[width=\textwidth]{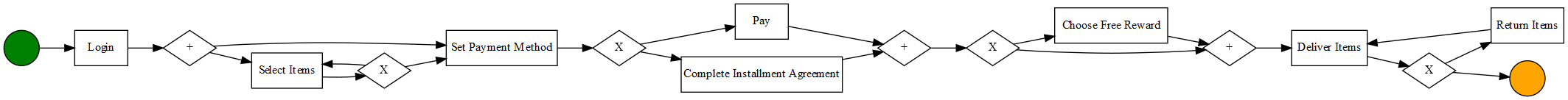}
        \caption{BPMN corresponding to the textual abstraction generated by GPT-4 using TA.\label{TA online shop}}
    \end{subfigure}
    \caption{BPMN models generated for the order handling process in the first run. Although the models generated using our system show some deviations from the original process description, the model generated by GPT-4 correctly captures complex non-hierarchical dependencies. Unlike the models generated using our system, TA led to an unsound model that is dead after the choice between paying and completing an installment agreement.} 
    \label{fig:order}
\end{figure}

\begin{figure}[!t]
    \centering   
    \begin{subfigure}[b]{\textwidth}
        \includegraphics[width=\textwidth]{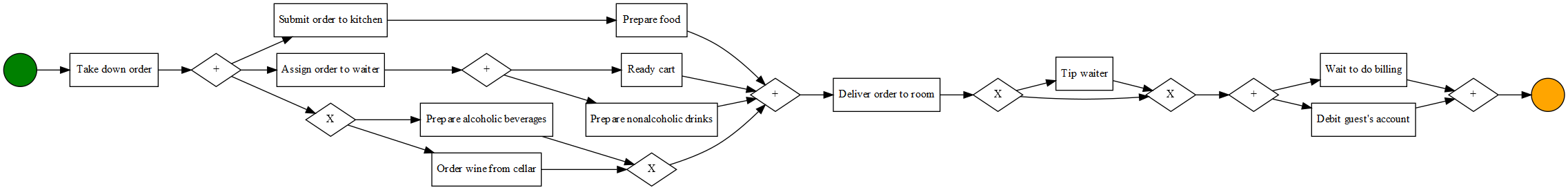}
        \caption{BPMN generated by GPT-4 using our system.\label{hotel_gpt4}}
    \end{subfigure}
    
    \begin{subfigure}[b]{\textwidth} 
        \includegraphics[width=\textwidth]{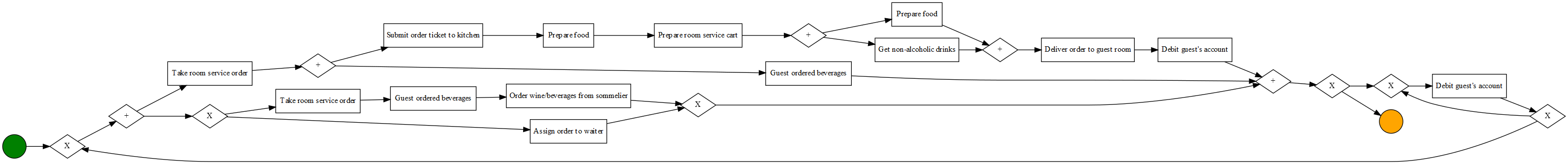}
        \caption{BPMN generated by Gemini using our system.}
    \end{subfigure}
    
    \begin{subfigure}[b]{\textwidth} 
        \includegraphics[width=\textwidth]{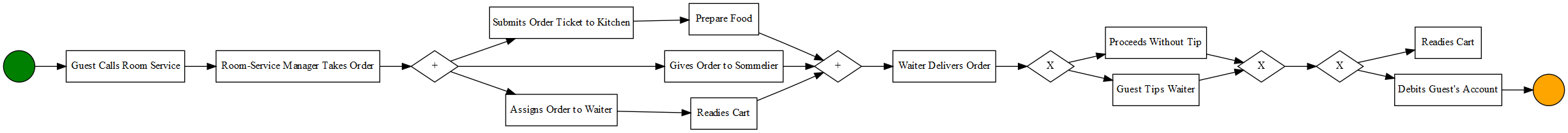}
        \caption{BPMN corresponding to the textual abstraction generated by GPT-4 using TA.\label{TA hotel}}
    \end{subfigure}
    \caption{BPMN models generated for the hotel process in the first run. The model generated by GPT-4 using our system provides a high degree of conformance with the process description, significantly surpassing the model generated by Gemini. The model generated using TA is unsound as the end event is not reachable after the second instance of ``Readies Cart''.} 
    \label{fig:hotel}
\end{figure}

In \autoref{tab:iterations_llm_grouped}, we report the number of error-handling iterations needed for the initial model generation and feedback integration for each process. We show the final models generated in the first run by both our framework and the TA framework: \autoref{fig:order} for the online shop process and \autoref{fig:hotel} for the hotel process. All generated models, along with an example detailing the complete sequence of prompts and responses exchanged between our system and GPT-4 until generating the final model, are available under \url{https://github.com/humam-kourani/LLM-Process-Modeling}. 

\subsection*{Addressing Q1}
GPT-4 demonstrated strong performance in generating process models for both processes. GPT-4 managed to deliver the initial models by the second error-handling iteration at the latest in all cases. Notably, the errors encountered during the generation of the model, which were classified as adjustable errors, were successfully resolved by GPT-4 in five cases. Feedback integration was notably efficient, with all feedback being accurately incorporated without any additional iterations for error handling in all three runs. 

The model shown in \autoref{hotel_gpt4} provides a high degree of conformance with the process description of the hotel process and the two employed feedback comments, however, some parts of the model can still be improved for better conformance. In the second run for the online shop process, GPT-4 was able to discover an optimal model that fully conforms with the reference model from \cite{powl}, showcasing its robust understanding and modeling capabilities. This process contains complex non-hierarchical dependencies between selecting the items, setting a payment method, the reward selection, and the payment choice. While conventional hierarchical process modeling languages, such as process trees, are unable to capture such complex dependencies, POWL empowers our framework with the capability to model these complex structures. The other models discovered by GPT-4 show some deviations from the original process. As LLMs continue to evolve, with ongoing advancements and enhancements, we expect future models to offer more consistency in the outcomes.

In contrast to GPT-4, Gemini's performance was significantly weaker. The quality of the models generated by Gemini is markedly inferior to those produced by GPT-4. Gemini struggled to properly resolve adjustable errors, and, although the initial model generation was successful in five of the six cases, this was due to the internal automatic error correction, not a resolution through the interaction with Gemini. Furthermore, Gemini failed to integrate the feedback comments in most cases, leading to the generation of critical errors. These errors included attempting to use non-existent functions, attempting to use external libraries, stopping the return of Python code, and ignoring instructions from the initial prompt. These issues highlight Gemini's limitations in understanding the task requirements and error resolution within our framework.

\subsection*{Addressing Q2}
Although some behaviors of the models produced by our framework deviate from the initial process descriptions, all produced models are sound and executable. The TA framework, in contrast, produces unsound models. For example, the model in \autoref{TA online shop} shows a choice between paying or completing an installment agreement through an exclusive choice gateway. The process is dead afterward; it requires both activities to be executed to proceed through the following parallel gateway. The model in \autoref{TA hotel} is also unsound as the end event is not reachable after the second instance of ``Readies Cart''. This shows the advantages of employing POWL as an intermediate process representation in ensuring the soundness of all models produced by our framework.

\subsection*{Evaluation Summary}
The comparative analysis between GPT-4 and Gemini demonstrates the superior capabilities of GPT-4 within our LLM-based process modeling framework. GPT-4 not only excelled in generating high-quality process models with remarkable efficiency but also showcased its adeptness at effectively resolving errors and seamlessly integrating user feedback. Our framework's comparison with the TA approach highlights its superiority, particularly in producing sound and executable models. This shows the robustness of our methodology and the strategic use of POWL as an intermediate process representation. 

\section{Limitations and Future Directions}\label{sec:limit}
Our approach, while pioneering in leveraging LLMs for process modeling, has limitations. In this section, we outline areas for improvement and propose ideas for addressing them in future work.

\paragraph{Expanding Process Perspectives.} Our framework addresses the control-flow perspective of process modeling, omitting the data, resource, and operational perspectives, which are crucial for a comprehensive understanding of business processes. The inherent flexibility and understanding capabilities of LLMs present a significant potential for extending our framework to incorporate additional process perspectives.

\paragraph{Extended Evaluation and User Studies.} While our evaluation demonstrates promising results with the datasets and process descriptions employed, we acknowledge the need for a broader investigation to better assess the generalizability of the framework. In our future work, we aim to extend the evaluation to encompass a more diverse set of processes and domains. Moreover, we aim to conduct a user study to evaluate the framework's usability, efficiency, and learning curve for both expert and non-expert users. 

\paragraph{Direct BPMN Generation.} The implemented system instantiating our framework utilizes POWL for intermediate process representation. A possible direction for future research is the exploration of the direct generation of BPMN models without an intermediate process representation. This approach promises to offer greater flexibility in representing intricate process structures and dynamics and allows for the enrichment of process models with context-rich annotations. However, moving away from the structured guarantees provided by POWL necessitates the development of more advanced process model generation and validation techniques. 

\paragraph{Enhanced Interactivity.} We intend to enhance the model refinement loop to support more nuanced and interactive feedback mechanisms. For example, we aim to empower users to not only provide textual feedback on generated process models but also to manually edit the generated models.

\section{Conclusion}
\label{sec:conclusion}
This paper introduces a novel framework that integrates LLMs with process modeling. Our framework leverages the natural language understanding and text generation capabilities of LLMs to generate and refine process models starting from textual descriptions. Our framework employs innovative prompting strategies for LLM utilization, a robust model generation protocol considering safety and quality aspects, and a user feedback mechanism for model refinement. While our framework enhances the accessibility and efficiency of process modeling, we recognize that manual effort remains crucial for validating generated models and providing effective feedback. Through preliminary results, we demonstrated the practicality and effectiveness of our framework, paving the way for future research and development.

\bibliographystyle{plain}
\bibliography{references}

\begin{thebibliography}{10}

\bibitem{DBLP:conf/aiia/BellanDG20}
Patrizio Bellan, Mauro Dragoni, and Chiara Ghidini.
\newblock A qualitative analysis of the state of the art in process extraction
  from text.
\newblock In Giuseppe Vizzari, Matteo Palmonari, and Andrea Orlandini, editors,
  {\em Proceedings of the AIxIA 2020 Discussion Papers Workshop co-located with
  the the 19th International Conference of the Italian Association for
  Artificial Intelligence (AIxIA2020), Anywhere, November 27th, 2020}, volume
  2776 of {\em {CEUR} Workshop Proceedings}, pages 19--30. CEUR-WS.org, 2020.

\bibitem{DBLP:conf/bpm/BellanADGP22}
Patrizio Bellan, Han van~der Aa, Mauro Dragoni, Chiara Ghidini, and
  Simone~Paolo Ponzetto.
\newblock {PET:} an annotated dataset for process extraction from natural
  language text tasks.
\newblock In Cristina Cabanillas, Niels~Frederik Garmann{-}Johnsen, and Agnes
  Koschmider, editors, {\em Business Process Management Workshops - {BPM} 2022
  International Workshops, M{\"{u}}nster, Germany, September 11-16, 2022,
  Revised Selected Papers}, volume 460 of {\em Lecture Notes in Business
  Information Processing}, pages 315--321. Springer, 2022.

\bibitem{DBLP:conf/bpm/Berti0A23}
Alessandro Berti, Daniel Schuster, and Wil M.~P. van~der Aalst.
\newblock Abstractions, scenarios, and prompt definitions for process mining
  with {LLMs}: {A} case study.
\newblock In Jochen~De Weerdt and Luise Pufahl, editors, {\em Business Process
  Management Workshops - {BPM} 2023 International Workshops, Utrecht, The
  Netherlands, September 11-15, 2023, Revised Selected Papers}, volume 492 of
  {\em Lecture Notes in Business Information Processing}, pages 427--439.
  Springer, 2023.

\bibitem{DBLP:conf/bpmds/BuschRSL23}
Kiran Busch, Alexander Rochlitzer, Diana Sola, and Henrik Leopold.
\newblock Just tell me: Prompt engineering in business process management.
\newblock In Han van~der Aa, Dominik Bork, Henderik~A. Proper, and Rainer
  Schmidt, editors, {\em Enterprise, Business-Process and Information Systems
  Modeling - 24th International Conference, {BPMDS} 2023, and 28th
  International Conference, {EMMSAD} 2023, Zaragoza, Spain, June 12-13, 2023,
  Proceedings}, volume 479 of {\em Lecture Notes in Business Information
  Processing}, pages 3--11. Springer, 2023.

\bibitem{DBLP:journals/aai/ChenL22b}
Song Chen and Hai Liao.
\newblock Bert-log: Anomaly detection for system logs based on pre-trained
  language model.
\newblock {\em Appl. Artif. Intell.}, 36(1), 2022.

\bibitem{DBLP:journals/jucs/GoncalvesSB11}
Jo{\~{a}}o~Carlos de~A.~R.~Gon{\c{c}}alves, Fl{\'{a}}via~Maria Santoro, and
  Fernanda~Ara{\'{u}}jo Bai{\~{a}}o.
\newblock Let me tell you a story - on how to build process models.
\newblock {\em J. Univers. Comput. Sci.}, 17(2):276--295, 2011.

\bibitem{DBLP:conf/naacl/DevlinCLT19}
Jacob Devlin, Ming{-}Wei Chang, Kenton Lee, and Kristina Toutanova.
\newblock {BERT:} pre-training of deep bidirectional transformers for language
  understanding.
\newblock In Jill Burstein, Christy Doran, and Thamar Solorio, editors, {\em
  Proceedings of the 2019 Conference of the North American Chapter of the
  Association for Computational Linguistics: Human Language Technologies,
  {NAACL-HLT} 2019, Minneapolis, MN, USA, June 2-7, 2019, Volume 1 (Long and
  Short Papers)}, pages 4171--4186. Association for Computational Linguistics,
  2019.

\bibitem{DBLP:journals/corr/abs-2312-11805}
Rohan~Anil et~al.
\newblock Gemini: {A} family of highly capable multimodal models.
\newblock {\em CoRR}, abs/2312.11805, 2023.

\bibitem{DBLP:conf/nips/BrownMRSKDNSSAA20}
Tom B.~Brown et~al.
\newblock Language models are few-shot learners.
\newblock In Hugo Larochelle, Marc'Aurelio Ranzato, Raia Hadsell,
  Maria{-}Florina Balcan, and Hsuan{-}Tien Lin, editors, {\em Advances in
  Neural Information Processing Systems 33: Annual Conference on Neural
  Information Processing Systems 2020, NeurIPS 2020, December 6-12, 2020,
  virtual}, 2020.

\bibitem{DBLP:journals/emisaij/FillFK23}
Hans{-}Georg Fill, Peter Fettke, and Julius K{\"{o}}pke.
\newblock Conceptual modeling and large language models: Impressions from first
  experiments with {ChatGPT}.
\newblock {\em Enterp. Model. Inf. Syst. Archit. Int. J. Concept. Model.},
  18:3, 2023.

\bibitem{DBLP:conf/caise/FriedrichMP11}
Fabian Friedrich, Jan Mendling, and Frank Puhlmann.
\newblock Process model generation from natural language text.
\newblock In Haralambos Mouratidis and Colette Rolland, editors, {\em Advanced
  Information Systems Engineering - 23rd International Conference, CAiSE 2011,
  London, UK, June 20-24, 2011. Proceedings}, volume 6741 of {\em Lecture Notes
  in Computer Science}, pages 482--496. Springer, 2011.

\bibitem{DBLP:conf/bpm/GrohsAER23}
Michael Grohs, Luka Abb, Nourhan Elsayed, and Jana{-}Rebecca Rehse.
\newblock Large language models can accomplish business process management
  tasks.
\newblock In Jochen~De Weerdt and Luise Pufahl, editors, {\em Business Process
  Management Workshops - {BPM} 2023 International Workshops, Utrecht, The
  Netherlands, September 11-15, 2023, Revised Selected Papers}, volume 492 of
  {\em Lecture Notes in Business Information Processing}, pages 453--465.
  Springer, 2023.

\bibitem{DBLP:conf/models/IvanchikjSP20}
Ana Ivanchikj, Souhaila Serbout, and Cesare Pautasso.
\newblock From text to visual {BPMN} process models: design and evaluation.
\newblock In Eugene Syriani, Houari~A. Sahraoui, Juan de~Lara, and Silvia
  Abrah{\~{a}}o, editors, {\em MoDELS '20: {ACM/IEEE} 23rd International
  Conference on Model Driven Engineering Languages and Systems, Virtual Event,
  Canada, 18-23 October, 2020}, pages 229--239. {ACM}, 2020.

\bibitem{DBLP:conf/bpm/KlievtsovaBKMR23}
Nataliia Klievtsova, Janik{-}Vasily Benzin, Timotheus Kampik, Juergen Mangler,
  and Stefanie Rinderle{-}Ma.
\newblock Conversational process modelling: State of the art, applications, and
  implications in practice.
\newblock In Chiara~Di Francescomarino, Andrea Burattin, Christian Janiesch,
  and Shazia~W. Sadiq, editors, {\em Business Process Management Forum - {BPM}
  2023 Forum, Utrecht, The Netherlands, September 11-15, 2023, Proceedings},
  volume 490 of {\em Lecture Notes in Business Information Processing}, pages
  319--336. Springer, 2023.

\bibitem{DBLP:conf/icpm/KouraniSA23}
Humam Kourani, Daniel Schuster, and Wil M.~P. van~der Aalst.
\newblock Scalable discovery of partially ordered workflow models with formal
  guarantees.
\newblock In {\em 5th International Conference on Process Mining, {ICPM} 2023,
  Rome, Italy, October 23-27, 2023}, pages 89--96. {IEEE}, 2023.

\bibitem{powl}
Humam Kourani and Sebastiaan~J. van Zelst.
\newblock {POWL:} partially ordered workflow language.
\newblock In Chiara~Di Francescomarino, Andrea Burattin, Christian Janiesch,
  and Shazia Sadiq, editors, {\em Business Process Management - 21st
  International Conference, {BPM} 2023, Utrecht, The Netherlands, September
  11-15, 2023, Proceedings}, volume 14159 of {\em Lecture Notes in Computer
  Science}, pages 92--108. Springer, 2023.

\bibitem{DBLP:series/lnbip/Leemans22}
Sander J.~J. Leemans.
\newblock {\em Robust Process Mining with Guarantees - Process Discovery,
  Conformance Checking and Enhancement}, volume 440 of {\em Lecture Notes in
  Business Information Processing}.
\newblock Springer, 2022.

\bibitem{ijcai2021p612}
Junyi Li, Tianyi Tang, Wayne~Xin Zhao, and Ji-Rong Wen.
\newblock Pretrained language model for text generation: A survey.
\newblock In Zhi-Hua Zhou, editor, {\em Proceedings of the Thirtieth
  International Joint Conference on Artificial Intelligence, {IJCAI-21}}, pages
  4492--4499. International Joint Conferences on Artificial Intelligence
  Organization, 8 2021.
\newblock Survey Track.

\bibitem{DBLP:conf/esws/MartinoIT23}
Ariana Martino, Michael Iannelli, and Coleen Truong.
\newblock Knowledge injection to counter large language model {(LLM)}
  hallucination.
\newblock In Catia Pesquita, Hala Skaf{-}Molli, Vasilis Efthymiou, Sabrina
  Kirrane, Axel Ngonga, Diego Collarana, Renato Cerqueira, Mehwish Alam,
  C{\'{a}}ssia Trojahn, and Sven Hertling, editors, {\em The Semantic Web:
  {ESWC} 2023 Satellite Events - Hersonissos, Crete, Greece, May 28 - June 1,
  2023, Proceedings}, volume 13998 of {\em Lecture Notes in Computer Science},
  pages 182--185. Springer, 2023.

\bibitem{DBLP:journals/corr/abs-2305-16807}
Daiki Miyake, Akihiro Iohara, Yu~Saito, and Toshiyuki Tanaka.
\newblock Negative-prompt inversion: Fast image inversion for editing with
  text-guided diffusion models.
\newblock {\em CoRR}, abs/2305.16807, 2023.

\bibitem{20.500.12116/43782}
Fabian Muff and Hans-Georg Fill.
\newblock Limitations of chatgpt in conceptual modeling: Insights from
  experiments in metamodeling, 2024.

\bibitem{DBLP:journals/corr/abs-2303-08774}
OpenAI.
\newblock {GPT-4} technical report.
\newblock {\em CoRR}, abs/2303.08774, 2023.

\bibitem{DBLP:journals/jksucis/SholiqSA22}
Sholiq Sholiq, Riyanarto Sarno, and Endang~Siti Astuti.
\newblock Generating {BPMN} diagram from textual requirements.
\newblock {\em J. King Saud Univ. Comput. Inf. Sci.}, 34(10 Part
  {B}):10079--10093, 2022.

\bibitem{DBLP:books/daglib/0027363}
Wil M.~P. van~der Aalst.
\newblock {\em Process Mining - Discovery, Conformance and Enhancement of
  Business Processes}.
\newblock Springer, 2011.

\bibitem{DBLP:journals/topnoc/HeeSW13a}
Kees~M. van Hee, Natalia Sidorova, and Jan Martijn E.~M. van~der Werf.
\newblock Business process modeling using {P}etri nets.
\newblock {\em Trans. {P}etri Nets Other Model. Concurr.}, 7:116--161, 2013.

\bibitem{DBLP:conf/hpec/VidanF23}
Andy Vidan and Lars~H. Fiedler.
\newblock A composable just-in-time programming framework with {LLMs} and
  {FBP}.
\newblock In {\em {IEEE} High Performance Extreme Computing Conference, {HPEC}
  2023, Boston, MA, USA, September 25-29, 2023}, pages 1--8. {IEEE}, 2023.

\bibitem{DBLP:conf/bpm/VidgofBM23}
Maxim Vidgof, Stefan Bachhofner, and Jan Mendling.
\newblock Large language models for business process management: Opportunities
  and challenges.
\newblock In Chiara~Di Francescomarino, Andrea Burattin, Christian Janiesch,
  and Shazia~W. Sadiq, editors, {\em Business Process Management Forum - {BPM}
  2023 Forum, Utrecht, The Netherlands, September 11-15, 2023, Proceedings},
  volume 490 of {\em Lecture Notes in Business Information Processing}, pages
  107--123. Springer, 2023.

\bibitem{DBLP:books/el/15/RosingWCM15}
Mark von Rosing, Stephen White, Fred Cummins, and Henk de~Man.
\newblock Business process model and notation - {BPMN}.
\newblock In Mark von Rosing, Henrik von Scheel, and August{-}Wilhelm Scheer,
  editors, {\em The Complete Business Process Handbook: Body of Knowledge from
  Process Modeling to BPM, Volume {I}}, pages 429--453. Morgan
  Kaufmann/Elsevier, 2015.

\bibitem{DBLP:journals/corr/abs-2305-14688}
Benfeng Xu, An~Yang, Junyang Lin, Quan Wang, Chang Zhou, Yongdong Zhang, and
  Zhendong Mao.
\newblock Expertprompting: Instructing large language models to be
  distinguished experts.
\newblock {\em CoRR}, abs/2305.14688, 2023.

\bibitem{DBLP:conf/iclr/ZhouMHPPCB23}
Yongchao Zhou, Andrei~Ioan Muresanu, Ziwen Han, Keiran Paster, Silviu Pitis,
  Harris Chan, and Jimmy Ba.
\newblock Large language models are human-level prompt engineers.
\newblock In {\em The Eleventh International Conference on Learning
  Representations, {ICLR} 2023, Kigali, Rwanda, May 1-5, 2023}. OpenReview.net,
  2023.

\end{thebibliography}

\end{document}